\documentclass[twocolumn]{aastex63}
\usepackage{natbib}
\usepackage{amsmath}
\usepackage{hyperref}
\usepackage[encapsulated]{CJK}

\newcommand{\msun}{$M_{\odot}$}

\newcommand{\mbh}{$M_{\rm BH}$}

\shorttitle{UNCOVER: UHZ-1 at $z = 10.1$}
\shortauthors{Goulding et al.}

\begin{document}

\title{UNCOVER: The growth of the first massive black holes from JWST/NIRSpec -- \\
spectroscopic redshift confirmation of an X-ray luminous AGN at z=10.1}

\correspondingauthor{Andy D. Goulding}
\email{goulding@astro.princeton.edu}

\author[0000-0003-4700-663X]{Andy D. Goulding}
\affiliation{Department of Astrophysical Sciences, Princeton University, Princeton, NJ 08544, USA}
\author[0000-0002-5612-3427]{Jenny E. Greene}
\affiliation{Department of Astrophysical Sciences, Princeton University, Princeton, NJ 08544, USA}
\author[0000-0003-4075-7393]{David J. Setton}
\affiliation{Department of Physics and Astronomy and PITT PACC, University of Pittsburgh, Pittsburgh, PA 15260, USA}
\author[0000-0002-2057-5376]{Ivo Labbe}
\affiliation{Centre for Astrophysics and Supercomputing, Swinburne University of Technology, Melbourne, VIC 3122, Australia}
\author[0000-0001-5063-8254]{Rachel Bezanson}
\affiliation{Department of Physics and Astronomy and PITT PACC, University of Pittsburgh, Pittsburgh, PA 15260, USA}
\author[0000-0001-8367-6265]{Tim B. Miller}
\affiliation{Department of Astronomy, Yale University, New Haven, CT 06511, USA}
\affiliation{Center for Interdisciplinary Exploration and Research in Astrophysics (CIERA) and
Department of Physics and Astronomy, Northwestern University, 1800 Sherman Ave, Evanston IL 60201, USA}

\author[0000-0002-7570-0824]{Hakim Atek}
\affiliation{Institut d'Astrophysique de Paris, CNRS, Sorbonne Universit\'e, 98bis Boulevard Arago, 75014, Paris, France}

\author[0000-0003-0573-7733]{\'Akos Bogd\'an}
\affil{Center for Astrophysics $\vert$ Harvard \& Smithsonian, 60 Garden Street, Cambridge, MA 02138, USA}

\author[0000-0003-2680-005X]{Gabriel Brammer}
\affiliation{Cosmic Dawn Center (DAWN), Niels Bohr Institute, University of Copenhagen, Jagtvej 128, K{\o}benhavn N, DK-2200, Denmark}

\author[0009-0009-9795-6167]{Iryna Chemerynska}
\affiliation{Institut d'Astrophysique de Paris, CNRS, Sorbonne Universit\'e, 98bis Boulevard Arago, 75014, Paris, France}

\author[0000-0002-7031-2865]{Sam E. Cutler}
\affiliation{Department of Astronomy, University of Massachusetts, Amherst, MA 01003, USA}

\author[0000-0001-8460-1564]{Pratika Dayal}\affil{Kapteyn Astronomical Institute, University of Groningen, 9700 AV Groningen, The Netherlands}

\author[0000-0001-7440-8832]{Yoshinobu Fudamoto}
\affiliation{Waseda Research Institute for Science and Engineering, Faculty of Science and Engineering, Waseda University, 3-4-1 Okubo, Shinjuku, Tokyo 169-8555, Japan}
\affiliation{National Astronomical Observatory of Japan, 2-21-1, Osawa, Mitaka, Tokyo, Japan}

\author[0000-0001-7201-5066]{Seiji Fujimoto}\altaffiliation{Hubble Fellow}
\affiliation{Department of Astronomy, The University of Texas at Austin, Austin, TX 78712, USA}

\author[0000-0001-6278-032X]{Lukas J. Furtak}\affiliation{Physics Department, Ben-Gurion University of the Negev, P.O. Box 653, Be’er-Sheva 84105, Israel}

\author[0000-0002-5588-9156]{Vasily Kokorev}
\affiliation{Kapteyn Astronomical Institute, University of Groningen, P.O. Box 800, 9700AV Groningen, The Netherlands}
\author[0000-0002-3475-7648]{Gourav Khullar}
\affiliation{Department of Physics and Astronomy and PITT PACC, University of Pittsburgh, Pittsburgh, PA 15260, USA}

\author[0000-0001-6755-1315]{Joel Leja}
\affiliation{Department of Astronomy \& Astrophysics, The Pennsylvania State University, University Park, PA 16802, USA}
\affiliation{Institute for Computational \& Data Sciences, The Pennsylvania State University, University Park, PA 16802, USA}
\affiliation{Institute for Gravitation and the Cosmos, The Pennsylvania State University, University Park, PA 16802, USA}

\author[0000-0001-9002-3502]{Danilo Marchesini}
\affiliation{Physics and Astronomy Department, Tufts University, 574 Boston Ave., Medford, MA 02155, USA}

\author[0000-0002-7809-0881]{Priyamvada Natarajan}
\affiliation{Department of Astronomy, Yale University, New Haven, CT 06511, USA}
\affiliation{Department of Physics, Yale University, New Haven, CT 06520, USA}
\affiliation{Black Hole Initiative, Harvard University, 20 Garden Street, Cambridge, MA 02138, USA}

\author[0000-0002-7524-374X]{Erica Nelson}\affiliation{Department for Astrophysical and Planetary Science, University of Colorado, Boulder, CO 80309, USA}

\author[0000-0001-5851-6649]{Pascal A. Oesch}
\affiliation{Department of Astronomy, University of Geneva, Chemin Pegasi 51, 1290 Versoix, Switzerland}
\affiliation{Cosmic Dawn Center (DAWN), Niels Bohr Institute, University of Copenhagen, Jagtvej 128, K{\o}benhavn N, DK-2200, Denmark}

\author[0000-0002-9651-5716]{Richard Pan}\affiliation{Department of Physics and Astronomy, Tufts University, 574 Boston Ave., Medford, MA 02155, USA}

\author[0000-0001-7503-8482]{Casey Papovich}
\affiliation{Department of Physics and Astronomy, Texas A\&M University, College
Station, TX, 77843-4242 USA}
\affiliation{George P.\ and Cynthia Woods Mitchell Institute for
 Fundamental Physics and Astronomy, Texas A\&M University, College
 Station, TX, 77843-4242 USA}

\author[0000-0002-0108-4176]{Sedona H. Price}
\affiliation{Department of Physics and Astronomy and PITT PACC, University of Pittsburgh, Pittsburgh, PA 15260, USA}

\author[0000-0002-8282-9888]{Pieter van Dokkum}
\affiliation{Department of Astronomy, Yale University, New Haven, CT 06511, USA}

\author[0000-0001-9269-5046]{Bingjie Wang (\begin{CJK*}{UTF8}{gbsn}王冰洁\ignorespacesafterend\end{CJK*})}
\affiliation{Department of Astronomy \& Astrophysics, The Pennsylvania State University, University Park, PA 16802, USA}
\affiliation{Institute for Computational \& Data Sciences, The Pennsylvania State University, University Park, PA 16802, USA}
\affiliation{Institute for Gravitation and the Cosmos, The Pennsylvania State University, University Park, PA 16802, USA}

\author[0000-0003-1614-196X]{John R. Weaver}
\affiliation{Department of Astronomy, University of Massachusetts, Amherst, MA 01003, USA}

\author[0000-0001-7160-3632]{Katherine E. Whitaker}
\affiliation{Department of Astronomy, University of Massachusetts, Amherst, MA 01003, USA}
\affiliation{Cosmic Dawn Center (DAWN), Denmark} 

\author[0000-0002-0350-4488]{Adi Zitrin}
\affiliation{Physics Department, Ben-Gurion University of the Negev, P.O. Box 653, Be’er-Sheva 84105, Israel}

\date{Aug 2023}
\submitjournal{ApJL}

\begin{abstract}
The James Webb Space Telescope is now detecting early black holes (BHs) as they transition from ``seeds'' to supermassive BHs. Recently, \citet{Bogdan:2023} reported the detection of an X-ray luminous supermassive BH, UHZ-1, with a photometric redshift at $z>10$. Such an extreme source at this very high redshift provides new insights on seeding and growth models for BHs given the short time available for formation and growth. Harnessing the exquisite sensitivity of JWST/NIRSpec, here we report the spectroscopic confirmation of UHZ-1 at $z=10.073 \pm 0.002$. We find that the NIRSpec/Prism spectrum is typical of recently discovered $z \approx 10$ galaxies, characterized primarily by star-formation features. We see no clear evidence of the powerful X-ray source in the rest-frame UV/optical spectrum, which may suggest heavy obscuration of the central BH, in line with the Compton-thick column density measured in the X-rays. We perform a stellar population fit simultaneously to the new NIRSpec spectroscopy and previously available photometry. The fit yields a stellar mass estimate for the host galaxy that is significantly better constrained than prior photometric estimates ($M_{\star} \sim 1.4^{+0.3}_{-0.4} \times 10^{8}$~\msun). Given the predicted BH mass ($M_{\rm BH} \sim 10^7-10^8$~\msun), the resulting ratio of $M_{\mathrm{BH}}$/$M_{\star}$ remains two to three orders of magnitude higher than local values, thus lending support to the heavy seeding channel for the formation of supermassive BHs within the first billion years of cosmic evolution.
\end{abstract}

\keywords{Active galactic nuclei (16), High-redshift galaxies (734), Early universe (435)}

\section{Introduction}

Until the launch of the	James Webb Space Telescope (JWST), the	earliest black holes known were a handful of extremely UV-luminous $z \approx 7$ quasars \citep[e.g.,][]{Mortlock:2011,Banados:2018,Matsuoka:2018,Matsuoka:2023}. While these sources are quite rare, given their high estimated masses, their existence leads to a timing challenge for the formation of supermassive black holes (SMBHs). If their initial seeds originate from the death of the first massive stars, with a typical remnant mass of $\sim 100$~\msun\, these black holes would need to accrete at or above the Eddington limit continuously for $\sim 700 - 800$ million years in order to reach the observed masses of $>10^9$~\msun\ \citep[see][for a review]{Natarajan:2011,Fan:2019}. Theorists have therefore explored alternate seed formation models, with heavier seed BH mass functions ($\sim 10^4$~\msun) that could form from the direct collapse of gas in the high redshift Universe \citep[see for instance]{lodatonatarajan2006,LodatoPN:2007, VolonteriGLPN:2008,Inayoshi:2022}. These heavy seeds are expected to be rare in general \citep[e.g.,][]{Agarwal:2013,Dayal:2019,Habouzit:2022}, and their individual future growth trajectories are unclear. Therefore, it is unclear if UHZ-1 is a likely progenitor for the luminous optically detected SDSS quasars. Guidance from cosmological simulations that track BH growth, the {\sc MASSIVEBLACK} suite in particular, have shown that the most massive BH at $z \sim 10$ does not necessarily grow to remain the most massive BH by $z=6$ \citep{DiMatteoetal:2017, DiMatteoetal:2023}. The details of the environment play an important role in shaping the accretion and, therefore, growth history of BHs.

The situation has gotten decidedly more	interesting with the launch of JWST. A number of intriguing active galactic nuclei (AGN) candidates have been spectroscopically confirmed at more moderate luminosities \citep{Kocevski:2023,Matthee:2023}, with some discovered at $z>7$ \citep{Harikane:2022,Furtak:2022,Larson:2023,Maiolino:2023a,Maiolino:2023b}. In the absence of other direct indicators of AGN activity (e.g., broadened Balmer emission lines), the most unambiguous identification of AGN activity is through the detection of high-energy X-ray emission. Due to the negative k-correction at X-ray energies, high redshift AGN are preferentially detected at increasingly harder X-ray energies, making their identification more distinct from lower energy X-rays produced by contaminating star-formation emission. 

Harnessing the exquisite spatial resolution afforded by the Chandra X-ray Observatory, the current record-holder to-date for the most distant ($z_{\rm phot} \sim 10.3$; \citealt{Castellano:2023}) X-ray luminous AGN is UHZ-1, discovered behind the lensing cluster Abell 2744 \citep[][]{Bogdan:2023}. Previously identified as an extremely high-redshift candidate \citep{Castellano:2023}, UHZ-1 is reported with a robust $4.2-4.4\sigma$ detection in the observed-frame 2--7~keV band with 20.6 net counts. At $z>10$, these are extremely hard X-ray photons with rest-frame energies of $E\sim22-80$~keV, which can only arise from an accreting BH. UHZ-1 is undetected in the softer 0.5--2~keV band, which the authors explain is likely due to UHZ-1 being heavily obscured with a Compton-thick column density of $N_{\rm H} \sim 10^{24} - 10^{25}$~cm$^{-2}$. Assuming their adopted lensing magnification of $\mu = 3.81$, and given the degeneracies related to the X-ray spectral fitting, the resultant intrinsic 2--10~keV X-ray luminosity is $L_{\rm X} \gtrsim 2 \times 10^{44}$~erg/s.

In this paper, we present the JWST/NIRSpec Prism spectroscopy confirming that UHZ1 is at a redshift of $z=10.07$ (\S~\ref{sec:data}). This deep spectrum was recently collected as part of the UNCOVER JWST treasury program (JWST-GO-2561, PIs: Labbe, Bezanson), which includes deep ($\sim2.7-17$ hr) low-resolution spectroscopy for $\sim700$ JWST-selected targets. We further examine the rest-frame UV/optical spectral properties of UHZ1 (\S~\ref{sec:properties}), and estimate the stellar mass and star-formation rate of the host galaxy (\S~\ref{sec:Bagpipes}). Throughout, we assume a $\Lambda$CDM cosmology with $\Omega_M = 0.29$, $\Omega_{\Lambda}=0.71$ and $H_0 = 69.6$ km s$^{-1}$ Mpc$^{-1}$, with a \citet{Chabrier:2003} initial mass function.
\begin{figure*}
\includegraphics[width=0.98\textwidth]{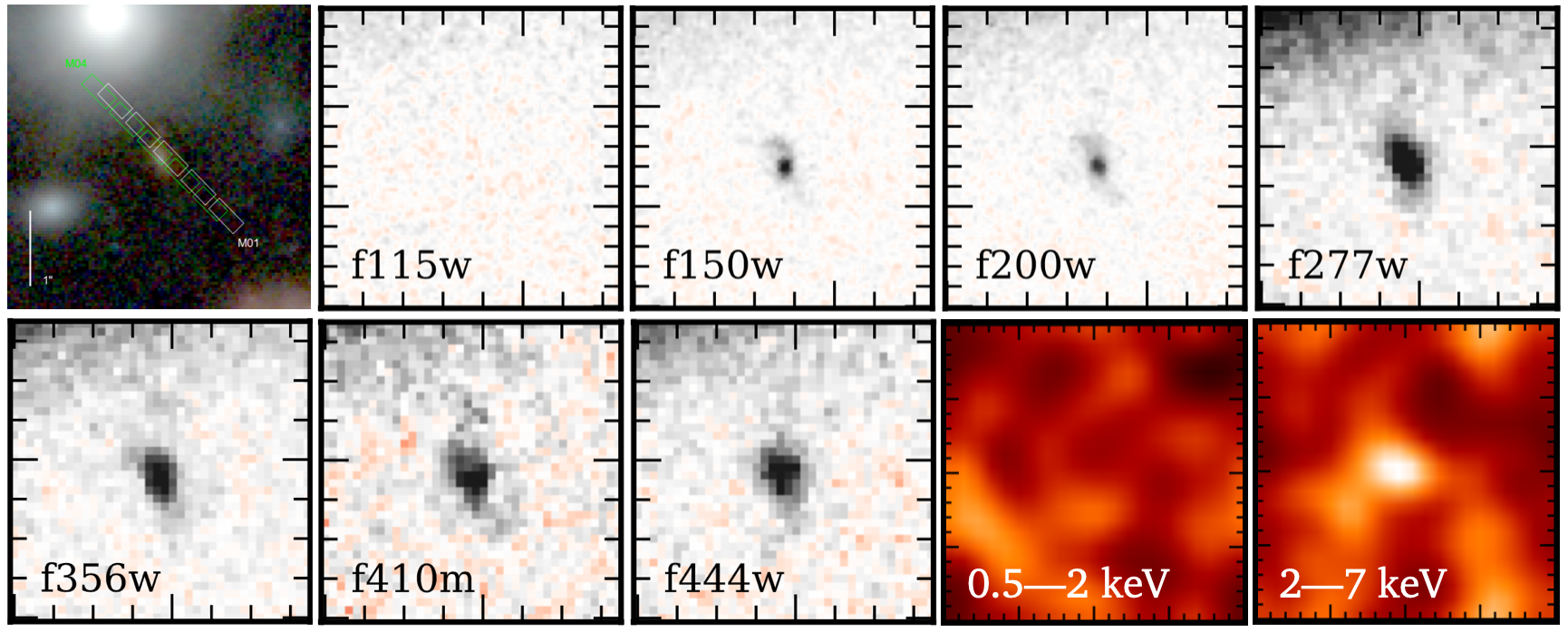}
\caption{From top-left: NIRSpec/Prism MSA shutter positions for UHZ1 (see \S~\ref{sec:data}), JWST/NIRCam images of UHZ1 in filters F115W, F150W, F200W, F277W, F356, F410M and F444W (photometric measurements from these calibrated data are presented in Fig.~\ref{fig:spectra}) and Chandra X-ray images in the 0.5-2 and 2-7~keV bands (smoothed with a 1~pixel width Gaussian filter). JWST cutout images are 1.5$''$ on a side, Chandra images are 5$''$ on a side. These are oriented in standard North-East convention.}
\label{fig:images}
\end{figure*}

\section{JWST/NIRSpec Prism Spectroscopy of UHZ1}
\label{sec:data}

\subsection{MSA Observational Setup}

UHZ-1 \citep[][]{Weaver:2023}, positioned at $\alpha=3.567070796^{\circ}$, $\delta=-30.3778606^{\circ}$, was observed on July 31 and August 1 2023 for a total of 7.1 hours as part of the Multi-shutter Array (MSA) follow-up program of the UNCOVER JWST field, Abell 2744 \citep{Bezanson:2022}. The NIRSpec/Prism observations of UNCOVER were split into 7 MSA configurations, with UHZ1 positioned on MSA 1 (2.7 hours) and 4 (4.4 hours) centered at $\alpha=3.5839128^{\circ}$, $\delta=-30.3998611^{\circ}$ and $\alpha=3.5586419^{\circ}$, $\delta=-30.3564066^{\circ}$, respectively. These observations employed a 2-POINT-WITH-NIRCam-SIZE2
dither pattern and a 3 shutter slit-let nod pattern at an angle of V3PA$\sim$266.0$^{\circ}$. MSA configuration 1 used the NRSIRS2RAPID readout pattern, while MSA 4 used NRSIRS2. For further details of the observational setup see \citet[][]{Bezanson:2022} and Price et al. (2023, in prep). Upon analysis it was determined that around the position of UHZ1, MSA 1 suffered from an electrical short, and given the resultant low signal to noise, we do not use those data for any part of the analysis presented here. Hence, the total usable exposure time for UHZ1 is 4.4 hours. 

\subsection{NIRSpec/Prism Data Reduction}

The Prism spectra are reduced using msaexp \citep[v0.6.10,][]{Brammer:22}. Beginning from the level 2 products downloaded from MAST\footnote{Available from: \dataset{http://dx.doi.org/10.17909/8k5c-xr27}}, 
\texttt{msaexp} applies a correction for 1/f noise, identifies and masks snowballs, and removes the bias in each individual frame. Parts of the JWST science calibration and reduction pipeline \citep{bushouse_howard_2023_7795697} are used to apply WCS, identify, flat-field, and apply photometric corrections to  each slit. 2D slits are then extracted and drizzled together onto a common grid. A local background subtraction is applied using vertically shifted, stacked 2D spectra. Finally, \texttt{msaexp} performs an optimal extraction using a Gaussian model to the collapsed spectrum with free center and width \citep[e.g.,][]{Horne:86}. This work is based on an early (internal v0.3) spectroscopic reduction. We perform slit-loss corrections by applying a wavelength-independent calibration to scale the normalization of the spectrum to the photometry by convolving the single-mask extracted 1D spectra with the broad/medium band filters, comparing to the total photometry (\citealt{Weaver:2023}), and modeling the wavelength dependent linear correction with a first order polynomial. Reduced data is planned for public release before Cycle 3 and presented in Price et al. (2023, in prep). 

\section{UHZ1: a z=10.07 X-ray luminous AGN}
\label{sec:properties}

\subsection{Prism spectroscopy}
\label{sec:spectra}

\begin{figure*}[t]
\centering
\includegraphics[width=0.96\textwidth]{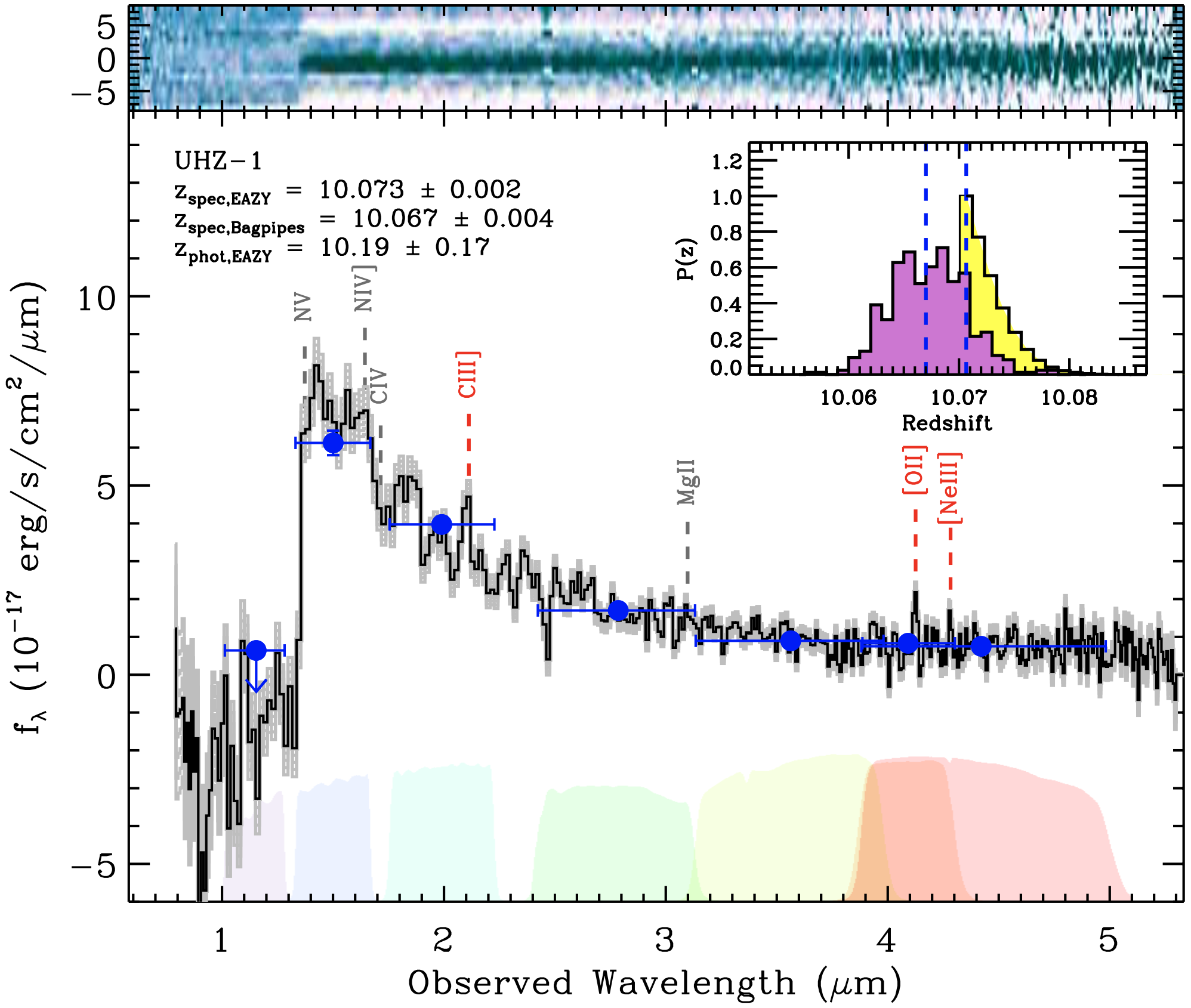}
\caption{JWST/NIRSpec Prism spectroscopy of UHZ-1. {\bf Upper panel:} 2D MSA Prism spectroscopy produced by {\tt msaexp}. {Lower panel:} 1D spectral extraction in $f_{\lambda}$ (in units of $10^{-17}$erg s$^{-1}$ cm$^{-2}$ $\mu$m$^{-1}$) with associated statistical uncertainties (gray shaded region). Slit-loss corrections are defined by convolution of the JWST photometry with the Prism spectrum (see \S~\ref{sec:data}). Prominent and/or expected emission features are highlighted assuming $z_{\rm spec}=10.07$ with significant $>3\sigma$ detections and non-detections labeled in red and gray, respectively. Overlaid are the JWST/NIRCam photometry (blue circles) with associated filter responses highlighted. {\bf Inset panel:} Redshift probability distributions for fits to the NIRSpec spectroscopy produced by {\tt EAZY} (yellow) and {\tt BAGPIPES} (purple) packages.} 
\label{fig:spectra}
\end{figure*}

The wide spectral coverage of the NIRSpec/Prism spectrum allows us to probe from the Lyman break all the way to $\sim 4500$~\AA\ rest-frame (Figure \ref{fig:spectra}). Using {\tt msaexp}, we fit stellar population and emission line templates to the spectroscopy, and unambiguously confirm the redshift $z=10.071 \pm 0.002$ with a strongly identified Lyman break and several emission lines. The previous best-fit photometric redshift solution for UHZ1 was $z_{\rm phot} \sim 10.19 \pm 0.17$ produced using the EAZY \citep{Brammer:2008} package (see \citealt{Atek:2023}), and $z_{\rm phot} \sim 10.87_{-0.41}^{+0.21}$ adapting the \texttt{Prospector}-$\beta$ model \citep{Johnson:2021,Wang:2023}. The inset of Fig.~\ref{fig:spectra} presents the redshift probability functions produced by {\tt msaexp} showing a narrow single-peaked P(z) at $z\sim10.07$, which is consistent with the previous best-fit photometric redshift. We further verify the high redshift nature of UHZ-1 using the publicly available Bayesian Analysis of Galaxies for Physical Inference and Parameter EStimation ({\tt Bagpipes}; \citealt{Carnall:2018,Carnall:2019}) code, finding a similar single-peaked redshift solution at $z=10.068\pm0.003$. In addition to the strong blue UV/optical continuum with slope $\beta \sim -2.7$ \citep{Castellano:2023}, we also note the presence of several weaker UV and optical lines, which we measure in the next section.

\subsection{Spectral Emission lines}

\begin{figure}
\includegraphics[width=0.95\linewidth]{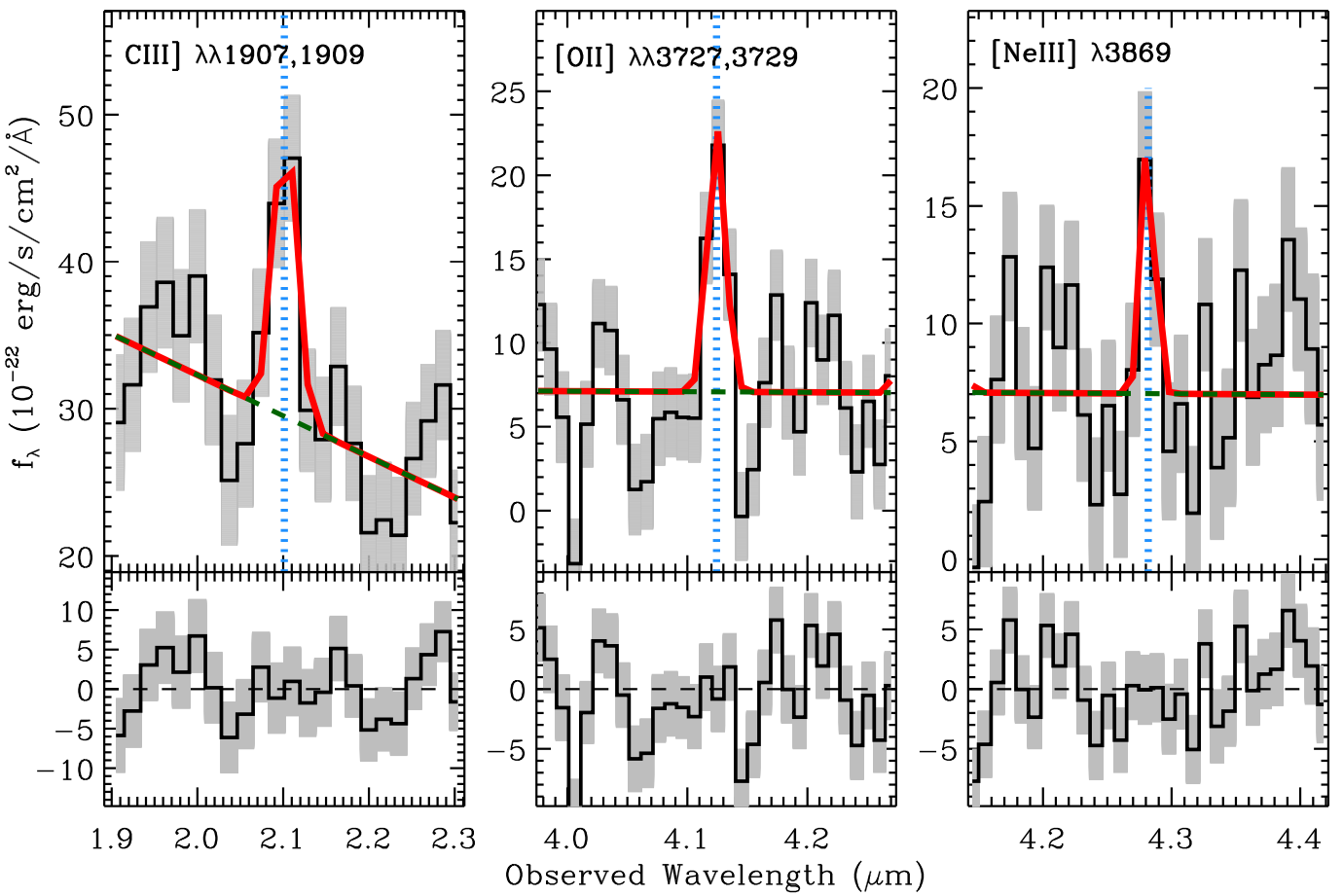}
\caption{{\bf Upper panels:} Gaussian line combined with low-order polynomial continua fits to the C{\sc iii}]~$\lambda \, 1907,1909$, [O{\sc ii}]~$\lambda \, 3727,3729$  doublet and [Ne{\sc iii}]~$\lambda \, 3869$ emission features. {\bf Lower panels:} Residuals of the best fits.}
\label{fig:emLines}
\end{figure}

We fit single Gaussians and low order polynomial continua to all prominent emission lines, and present the emission line centroids, fluxes and signal to noise ratios in Table~\ref{tab:emission}. We find significant detections of C{\sc iii}]~$\lambda \, 1907,1909$, [Ne{\sc iii}]~$\lambda \, 3869$, and the [O{\sc ii}]~$\lambda \, 3727,3729$ doublet (see Fig.~\ref{fig:emLines}). Interestingly, despite the clear and robust X-ray detection of UHZ1, and hence its identification as an AGN, we do not detect evidence for any broadened emission lines such as C{\sc iv} or MgII as expected for relatively unobscured AGN. Even for obscured AGN, typically high-EW, high-ionization narrow emission lines such as [Ne{\sc v}]$\lambda$3426 are detected, which would point to the presence of a rapidly accreting BH. Instead, the only clear signature of AGN activity in this source comes from the strong detection of X-ray emission at hard energies. While lower ionization lines such as [Ne{\sc iii}] can indicate the presence of an AGN, they are also readily produced by star formation, and this alone would not be the basis for an unambiguous AGN indicator (e.g., \citealt{Goulding:2009}). The UV luminosity ($M_{\rm UV} \sim -19.85$) is also not distinctively high compared to similarly high-z star-forming galaxies. Taken together, the lack of obvious AGN emission lines and the clear detection of luminous hard X-ray emission, we can conclude that UHZ-1 is likely an extremely Compton-thick and optically-obscured AGN, potentially similar to systems such as NGC~4945 in the local Universe \citep{Matt:2000,Yaqoob:2012}.

\begin{table}[t!]
\begin{center}
\setlength{\tabcolsep}{1.5mm}
\caption{Emission Line Measurements of UHZ1}
\begin{tabular}{lccc}

\hline \hline

\multicolumn{1}{c}{Name}  & 
\multicolumn{1}{c}{Obs. Cent.} & 
\multicolumn{1}{c}{Flux}  & 
\multicolumn{1}{c}{S/N} \\
\multicolumn{1}{c}{}  & 
\multicolumn{1}{c}{ $\lambda$ ($\mu m$)}  & 
\multicolumn{1}{c}{($10^{-20}$ erg s$^{-1}$ cm$^{-2}$)}  & 
\multicolumn{1}{c}{} \\

\hline
N\sc{iv}] $\lambda1485$ & 1.637 & $<112.40$ & $<3$ \\	
C\sc{iv} $\lambda\lambda1548,1550$ & 1.715 & $<99.04$ & $<3$ \\	
C\sc{iii}] $\lambda\lambda1907,1909$ & 2.101 & $67.93 \pm 17.62$ & 3.9 \\	
Mg\sc{ii} $\lambda2799$ & 3.098 & $< 30.03$ & $<3$ \\	
$[$O\sc{ii}] $\lambda\lambda 3727,3729$ & 4.124 & $29.29 \pm 5.50$ & 5.3 \\
$[$Ne\sc{iii}] $\lambda\lambda 3869$ & 4.282 & $14.63 \pm 4.77$ & 3.1 \\
\hline \hline
\end{tabular}
\label{tab:emission}
\end{center}

\normalsize
\end{table}

\section{Host Galaxy Properties}
\label{sec:Bagpipes}

Based upon the non-detection of broadened emission lines as well as lack of prominent high-ionization lines and a blue UV continuum shape that is consistent with star formation, there is no clear and direct evidence for a significant contribution from the AGN to the UV/optical spectrum. Therefore, we can fit the spectrophotometry robustly, where the light contribution is dominated by the host galaxy, despite knowledge of the presence of an X-ray AGN.

We use the \texttt{Bagpipes} SED fitting code \citep{Carnall:2018,Carnall:2019} to perform this SED fit. We use \cite{Bruzual:2003} stellar population models, the MILES spectral library \citep{Sanchez-Bazquez:2006,Falcon-Barroso2011}, a \citet{Chabrier:2003} IMF, and Cloudy nebular emission models \citep{Ferland:2017}. We utilize \texttt{PyMultinest} \citep{Buchner:2014, Feroz:2019} to perform our sampling using the default \texttt{Bagpipes} convergence criteria. We parameterize the star formation history with a delayed-$\tau$ model (SFR $\propto \ t e^{-t/ \tau}$) with the age and $\tau$ as free parameters ($0.1 \ \mathrm{Gyr}<\mathrm{age} <t_\mathrm{universe}$ and $0.01 \ \mathrm{Gyr}<\tau<5 \ \mathrm{Gyr}$). We additionally allow the metallicity (with the stellar and gas phase metallicity fixed to the same value) and ionization parameter to vary, with $-2<\mathrm{log(Z)}<0.3$ and $-3.5<\mathrm{log(U)}<-1.0$. We assume a \cite{Charlot:2000} dust model, with $0<A_v<5$ and $0.3<n<2.5$ as free parameters. We allow the redshift to vary around the best fitting spectroscopic redshift of 10.07$\pm$0.05. Additionally, we apply the wavelength-dependent instrumental resolution curve to all models before fitting, assuming that the resolution is a nominal 1.3 times better than the pre-flight curve provided by STSci\footnote{https://jwst-docs.stsci.edu/jwst-near-infrared-spectrograph/nirspec-instrumentation/nirspec-dispersers-and-filters} \citep{Curtis-Lake:2023}, and we leave the wavelength-independent velocity dispersion free between 1 km/s and 2000 km/s as a nuisance parameter. We fit for a polynomial calibration vector of order 2 and the \texttt{Bagpipes} white noise model to allow for underestimated errors up to a factor of 10. Finally, we place a signal-to-noise ceiling of 20 on both our photometry and spectroscopy to account for potential systematic issues with the flux calibration. We fit all available \textit{JWST}/NIRCam (F115W/F150W/F200W/F277W/F356W/F410M/F444W) and \textit{HST}/ACS (F435W/F606W/F814W) and WFC3 (F105W/F125W/F160W) photometry in addition to the full NIRSPEC/Prism spectrum.

\begin{figure*}
\includegraphics[width=1.0\textwidth]{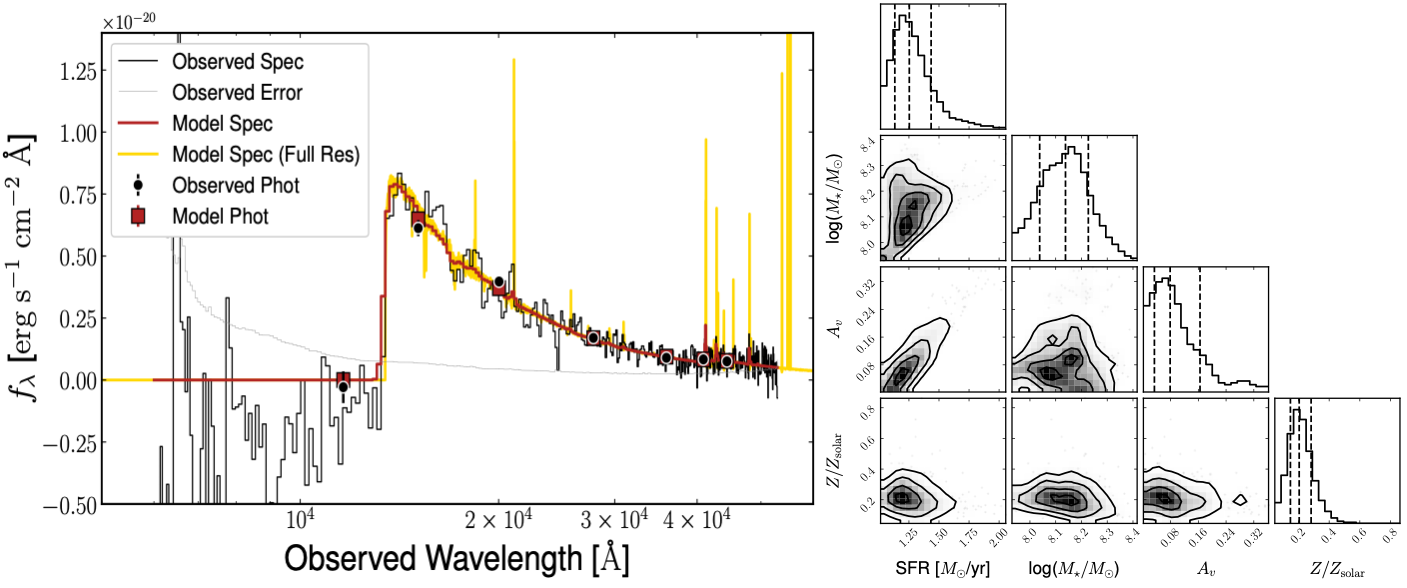}

\caption{SPS modeling with \texttt{Bagpipes}. The left panel shows the observed galaxy SED and spectrum (black, showing only JWST photometry) after the application of the polynomial calibration with 1$\sigma$ errors (assuming an error floor of 5\%) and the median model (red). 
Additionally, the median model is shown at the full resolution (gold), highlighting the predicted emission features that are washed out by the instrumental resolution. 
In the right panel, we show the covariant posteriors for a number of key measured parameters after accounting for the magnification of the source.}
\label{fig:bagpipes}
\end{figure*}

We present the best-fit model spectrum in Figure~\ref{fig:bagpipes}a along with the pairwise posterior distributions of relevant parameters in Figure~\ref{fig:bagpipes}b, which have been corrected for the magnification of UHZ1 with the median value of $\mu = 3.866$ \citep[$\pm^{0.1}_{0.4}$, see][]{Furtak:2023b}. The median, 16th and 84th percentiles of the host galaxy parameter posteriors are presented in Table~\ref{tab:host}. The magnification-corrected stellar mass of log$(M_\star/M_\odot) = 8.1 \pm 0.1$ is fairly typical for $z \approx 10$ galaxies being spectroscopically confirmed with \emph{JWST} \citep[e.g.,][]{Curtis-Lake:2023,Arrabal-Haro:2023}, but is now more robustly constrained to the higher mass end of the stellar mass previously inferred from photometry alone \citep{Castellano:2022,Bogdan:2023} owing to the precise nature of the spectroscopic redshift in the {\tt Bagpipes} fit. We note that fitting only the photometric data at the spectroscopic redshift of UHZ-1 produces a consistent measure of the stellar mass with log$(M_\star/M_\odot) = 8.1 \pm 0.1$. The {\tt Bagpipes} fit prefers a star-formation rate of SFR$\sim 1.3 \pm 0.2$~\msun/yr, i.e., a sSFR$\sim 10^{-8}$~yr$^{-1}$, which is similar to within a factor $\sim 2-3$ of sSFR values inferred for galaxies of similar mass at this early epoch \citep[e.g.,][]{Castellano:2023}. Finally, the fit prefers a moderately low stellar metallicity of $Z = 0.2 \pm 0.05$ compared to solar metallicity, similar to other early galaxies.  

We perform a parametric Sersic-profile fit to the NIRCam F444W image using \texttt{pysersic} \citep{Pasha:2023}, utilizing the NUTS sampler to explore the posterior \citep{Hoffman2014,Phan2019}. We find the Sersic index is consistent with $n\sim1$, typical of disk-like structures, and with an effective radius of $r_{\rm eff} \sim 0.14 \pm 0.02$ arcseconds, which at $z=10.07$ translates to a physical scale of $r_{\rm eff,physical}\sim 0.592$~kpc. We therefore find that UHZ-1 is also fairly typical in physical scales and extent measured for high-z galaxies in the GLASS field \citep{Castellano:2022}.

\begin{table}
\setlength{\tabcolsep}{1.8mm}
\caption{Host Galaxy Parameters}
\begin{tabular}{lcccc}

\hline \hline

\multicolumn{1}{c}{Parameter}  & 
\multicolumn{1}{c}{Unit} & 
\multicolumn{3}{c}{Percentile} \\
\multicolumn{1}{c}{}  & 
\multicolumn{1}{c}{}  & 
\multicolumn{1}{c}{16th}  & 
\multicolumn{1}{c}{50th} &
\multicolumn{1}{c}{84th}  \\ 

\hline
SFR & $M_{\odot}$/yr & 1.13 & 1.25 & 1.43 \\	
log($M_{\star}/M_{\odot}$) &  & 8.03 & 8.14 & 8.23 \\	
logU & - & -2.00 & -1.78 & -1.58 \\	
$A_{\rm V}$ & mag & 0.03 & 0.08 & 0.16 \\	
$t_{50}$ & Myr & 45.1 & 64.9 & 96.6 \\	
$Z$ & $Z_{\odot}$ & 0.15 & 0.20 & 0.28 \\	
$r_{\rm eff}$ & arcsec & 0.12 & 0.14 & 0.16 \\	
Sersic $n$ & - & 0.79 & 1.06 & 1.64 \\	

\hline \hline
\end{tabular}
\label{tab:host}

\normalsize
\end{table}

\section{Discussion and Summary}
\label{sec:discussion}

We have performed JWST NIRSpec/Prism spectroscopic follow-up of the $z\approx 10$ X-ray luminous AGN, UHZ-1, and confirm the highest-redshift (currently known) galaxy hosting an X-ray AGN, with a firm spectroscopic redshift of $z=10.073 \pm 0.002$. The spectrum is likely dominated by host galaxy light in the rest-frame UV/optical and stellar population synthesis analysis yields a stellar mass for the system of log $M_* = 8.1 \pm 0.1$~\msun. The size, mass, emission-line properties and star-formation rate of the host of UHZ-1 seem relatively typical of other known $z \approx 10$ galaxies. Here, we discuss the implications of UHZ1 for the growth of supermassive BHs, and for their signatures in the very early Universe.

A large open question for the formation of supermassive black holes has been whether they originate from ``light'' or stellar-mass black holes, remnants from the death of massive stars \citep[$\sim 100$~\msun\ seeds;][]{loebrasio1994,brommloeb2003} or whether there are mechanisms that operate to form heavier initial seeds \citep[\mbh$\sim 10^3-10^5$~\msun;][]{millerhamilton2002,portegieszwartetal2002,freitagetal2006,devecchivolonteri2009, koushiappasetal2004,lodatonatarajan2006,LodatoPN:2007,TalPN:2014,begelman2010,natarajanetal2017,Natarajan:2021,Greeneetal:2020}. As shown by \citet{Bogdan:2023}, to form the BH in UHZ-1 requires either continuous growth exceeding the Eddington limit for $>200$ Myr, or a massive seed. Similar timing arguments have been made for UV-luminous quasars at $z>6$; while they have more time to grow, they pose challenges for our current understanding of accretion models \citep[e.g.,][]{Haiman:2001,Natarajan:2011}.

The ratio of BH mass to galaxy mass provides a complementary clue to the seeding mechanism. A key prediction of the heavy seed formation scenarios by the direct collapse of gas is that this ratio at early times, close to the seeding epoch, is significantly higher than found locally \citep{Natarajan:2011,natarajanetal2017}. Heavy seeding models predict an early phase of overly massive BHs relative to the local $M_{\rm BH}/M_*$ relation \citep[e.g.,][]{natarajanetal2017}. Under the assumption that the AGN is contributing little to the observed UV/optical continuum and emission lines, the measured host mass for UHZ-1 is roughly 0.5 dex higher with the benefit of spectroscopy compared to the median photometric value in \citet{Castellano:2023}. However, we still estimate a very high ratio of $M_{\rm BH}/M_*$ based on the X-ray emission. The X-ray source in UHZ-1 has an $L_{\rm X, int} \sim 2 \times 10^{44}$~erg/s (assuming $N_{\rm H}\sim 2 \times 10^{24}$~cm$^{-2}$), and an implied BH mass of $M_{\rm BH} \sim 10^{7-8}$~\msun\ conservatively assuming Eddington limited accretion and relevant uncertainties regarding the high Compton-thick column density observed in the X-rays \citep{Bogdan:2023}. The implied BH to stellar mass ratio in this system (combining relevant modeling uncertainties) based on our best-fit {\tt Bagpipes} model is thus $M_{\rm BH}/M_{\rm gal} \approx 0.05-1.0$. 

This BH is a much larger fraction of the galaxy mass than the typical ratio of $\sim 0.001-0.002$ seen locally \citep{kormendyho2013,reinesvolonteri2015}. At face value, this observation would strongly disfavor light seed scenarios (in systems similar to UHZ-1) which typically result in BHs that are under massive with respect to the host galaxy until the galaxy mass exceeds $\sim{}10^{10}\,M_\odot$ \cite{AnglesAlcazar:2017,Catmabacak:2022}. Thus far, all luminous AGN discovered at $z>5$ with estimates for both the BH mass and the galaxy mass have had high ratios, but this source is extreme even in that context \citep[e.g.,][]{Izumi:2019,Neeleman:2021,Fan:2023}. This of course does not preclude a light-seed scenario concurrently taking place in other systems, particularly as such lower-mass BHs would be substantially more difficult to detect and may not be X-ray luminous \citep{Natarajan:2021}. However, the combination of the high BH mass, low galaxy mass at $z \sim 10$ and early accretion history modeling suggest that the particular BH in UHZ-1 was likely formed from a heavy seed \citep{Natarajan:2023}. Indeed, given the extremely small volume probed by the UNCOVER region ($\sim$40~arcmin$^2$), the number density of BHs formed from a heavy seed or a scenario involving light seeds that require substantial super-Eddington growth, cannot be vanishingly small with the detection of UHZ-1 and other systems thus far.

Despite the $> 4\sigma$ detection in the 2--7~keV band \citep[see Fig.~\ref{fig:images} and][]{Bogdan:2023}, the lack of clear AGN signatures in the UV/optical given the luminous hard X-ray source present in this system is somewhat surprising. While highly unlikely given the precise locale and centroiding of the X-ray emission, it cannot be excluded that the X-ray association is coincidental. On the other hand, the lack of UV AGN signatures is not unusual. Even in other bonafide broad-line AGN detected by JWST, which interestingly lack X-ray emission despite their Type-1 AGN nature (e.g., \citealt{Furtak_nature:2023}), these systems are also bereft of the typical broad and narrow AGN emission lines blueward of $H\beta$ (e.g., \citealt{Harikane:2023,Kocevski:2023,Matthee:2023, LabbeGreene:2023}). Moreover, such objects are not entirely uncommon in the nearby Universe (e.g., NGC 1448; NGC 4945; see \citealt{Alexander:2012,Annuar:2017}) as even luminous AGN signatures can be swamped by strong star formation and/or extreme obscuration. Furthermore, the relatively high $L_{\rm X,int} / L_{\rm 1450,obs} \sim 0.7$ ratio also points towards a heavily buried AGN. Indeed, the relatively low $A_{\rm V} \sim 0.08$~mag suggests a low line-of-sight extinction to the galaxy, placing any obscuration on small nuclear scales, as suggested by the Compton-thick column measured in the X-rays. Longer wavelength data may provide useful clues about the AGN's nature in UHZ-1. For example, due to the spectral coverage of the NIRSpec/Prism data, the current spectrum just misses the vital [O{\sc III}] and H$\beta$ complex (a typical diagnostic for AGN activity). Moreover, at significantly longer wavelengths, the rest-frame near and mid-IR wavebands can provide much cleaner information on the presence of AGN emission even in the most heavily obscured AGN due to the re-emission by dust of the AGN signatures. Such wavelengths are fully accessible by JWST/MIRI out to restframe $\lambda < 2.5 \mu$m, giving access to the hot AGN dust continuum as well as high-ionization emission lines and the Paschen series. Indeed, the detection of lower-order Balmer lines and the Paschen series, given the lack of Balmer emission lines in the current NIRSpec/Prism data may point towards heavy obscuration in UHZ-1.

Objects such as UHZ-1 are only now beginning to be uncovered in the JWST data. The UNCOVER program hints at the power of deep NIRSpec spectroscopy to characterize the first growing black holes, while the high magnification ($\mu \sim 3.8 \pm 0.2$) certainly aids in the detection of objects such as that presented here. The detection of the unimpeded UV-photons produced due to star formation from UHZ-1 suggests that the obscuration of the growing BH is confined to the nuclear regions given the low inferred extinction along the line of sight. This suggests that at extremely high-redshift, even modest star-forming sources like UHZ-1, likely jump-started re-ionization. The number of AGN detected so far, given the relatively limited areas covered by JWST to date, suggests that an actively growing BH population may already be in place at this early epoch. However, because of the difficulty to detect the AGN signatures in UHZ-1, the further identification and characterization of the demographics of such systems may require a suite of diagnostics from JWST NIRCam, NIRSpec, MIRI and in the X-rays.

\section*{Acknowledgments}
The authors thank the anonymous referee for a constructive and insightful report that improved several aspects of the manuscript. A.D.G would like to thank M.A.~Strauss for useful and enlightening conversations. A.D.G and J.E.G. acknowledges support from NSF/AAG grant\# 1007094. PD acknowledges support from the NWO grant 016.VIDI.189.162 (``ODIN") and the European Commission's and University of Groningen's CO-FUND Rosalind Franklin program. RB acknowledges support from the Research Corporation for Scientific Advancement (RCSA) Cottrell Scholar Award ID No: 27587. This work is based in part on observations made with the NASA/ESA/CSA James Webb Space Telescope. The data were obtained from the Mikulski Archive for Space Telescopes at the Space Telescope Science Institute, which is operated by the Association of Universities for Research in Astronomy, Inc., under NASA contract NAS 5-03127 for JWST. The specific observations analyzed can be accessed via \dataset{http://dx.doi.org/10.17909/8k5c-xr27}. These observations are associated with JWST Cycle 1 programs JWST-GO-2561 and JWST-ERS-1324. Support for program JWST-GO-2561 was provided by NASA through a grant from the Space Telescope Science Institute, which is operated by the Associations of Universities for Research in Astronomy, Incorporated, under NASA contract NAS5-26555. The Cosmic Dawn Center is funded by the Danish National Research Foundation (DNRF) under grant \#140. YF acknowledge support from NAOJ ALMA Scientific Research Grant number 2020-16B. YF further acknowledges support from support from JSPS KAKENHI Grant Number JP23K13149. HA and IC acknowledge support from CNES, focused on the JWST mission, and the Programme National Cosmology and Galaxies (PNCG) of CNRS/INSU with INP and IN2P3, co-funded by CEA and CNES.  \'AB acknowledges support from the Smithsonian Institution through NASA contract NAS8-03060. PN acknowledges support from the Gordon and Betty Moore Foundation and the John Templeton Foundation that fund the Black Hole Initiative (BHI) at Harvard University where she serves as one of the PIs. AZ acknowledges support by Grant No. 2020750 from the United States-Israel Binational Science Foundation (BSF) and Grant No. 2109066 from the United States National Science Foundation (NSF), and by the Ministry of Science \& Technology, Israel.

\bibliographystyle{aasjournal}
\bibliography{imbh.bib}

\begin{thebibliography}{}
\expandafter\ifx\csname natexlab\endcsname\relax\def\natexlab#1{#1}\fi
\providecommand{\url}[1]{\href{#1}{#1}}
\providecommand{\dodoi}[1]{doi:~\href{http://doi.org/#1}{\nolinkurl{#1}}}
\providecommand{\doeprint}[1]{\href{http://ascl.net/#1}{\nolinkurl{http://ascl.net/#1}}}
\providecommand{\doarXiv}[1]{\href{https://arxiv.org/abs/#1}{\nolinkurl{https://arxiv.org/abs/#1}}}

\bibitem[{{Agarwal} {et~al.}(2013){Agarwal}, {Davis}, {Khochfar}, {Natarajan},
  \& {Dunlop}}]{Agarwal:2013}
{Agarwal}, B., {Davis}, A.~J., {Khochfar}, S., {Natarajan}, P., \& {Dunlop},
  J.~S. 2013, \mnras, 432, 3438, \dodoi{10.1093/mnras/stt696}

\bibitem[{{Alexander} \& {Hickox}(2012)}]{Alexander:2012}
{Alexander}, D.~M., \& {Hickox}, R.~C. 2012, \nar, 56, 93,
  \dodoi{10.1016/j.newar.2011.11.003}

\bibitem[{{Alexander} \& {Natarajan}(2014)}]{TalPN:2014}
{Alexander}, T., \& {Natarajan}, P. 2014, Science, 345, 1330,
  \dodoi{10.1126/science.1251053}

\bibitem[{{Angl{\'e}s-Alc{\'a}zar} {et~al.}(2017){Angl{\'e}s-Alc{\'a}zar},
  {Faucher-Gigu{\`e}re}, {Quataert}, {Hopkins}, {Feldmann}, {Torrey}, {Wetzel},
  \& {Kere{\v{s}}}}]{AnglesAlcazar:2017}
{Angl{\'e}s-Alc{\'a}zar}, D., {Faucher-Gigu{\`e}re}, C.-A., {Quataert}, E.,
  {et~al.} 2017, \mnras, 472, L109, \dodoi{10.1093/mnrasl/slx161}

\bibitem[{{Annuar} {et~al.}(2017){Annuar}, {Alexander}, {Gandhi}, {Lansbury},
  {Asmus}, {Ballantyne}, {Bauer}, {Boggs}, {Boorman}, {Brandt}, {Brightman},
  {Christensen}, {Craig}, {Farrah}, {Goulding}, {Hailey}, {Harrison}, {Koss},
  {LaMassa}, {Murray}, {Ricci}, {Rosario}, {Stanley}, {Stern}, \&
  {Zhang}}]{Annuar:2017}
{Annuar}, A., {Alexander}, D.~M., {Gandhi}, P., {et~al.} 2017, \apj, 836, 165,
  \dodoi{10.3847/1538-4357/836/2/165}

\bibitem[{{Arrabal Haro} {et~al.}(2023){Arrabal Haro}, {Dickinson},
  {Finkelstein}, {Fujimoto}, {Fern{\'a}ndez}, {Kartaltepe}, {Jung}, {Cole},
  {Burgarella}, {Chworowsky}, {Hutchison}, {Morales}, {Papovich}, {Simons},
  {Amor{\'\i}n}, {Backhaus}, {Bagley}, {Bisigello}, {Calabr{\`o}},
  {Castellano}, {Cleri}, {Dav{\'e}}, {Dekel}, {Ferguson}, {Fontana}, {Gawiser},
  {Giavalisco}, {Harish}, {Hathi}, {Hirschmann}, {Holwerda}, {Huertas-Company},
  {Koekemoer}, {Larson}, {Lucas}, {Mobasher}, {P{\'e}rez-Gonz{\'a}lez},
  {Pirzkal}, {Rose}, {Santini}, {Trump}, {de la Vega}, {Wang}, {Weiner},
  {Wilkins}, {Yang}, {Yung}, \& {Zavala}}]{Arrabal-Haro:2023}
{Arrabal Haro}, P., {Dickinson}, M., {Finkelstein}, S.~L., {et~al.} 2023,
  \apjl, 951, L22, \dodoi{10.3847/2041-8213/acdd54}

\bibitem[{{Atek} {et~al.}(2023){Atek}, {Chemerynska}, {Wang}, {Furtak},
  {Weibel}, {Oesch}, {Weaver}, {Labb{\'e}}, {Bezanson}, {van Dokkum}, {Zitrin},
  {Dayal}, {Williams}, {Nannayakkara}, {Price}, {Brammer}, {Goulding}, {Leja},
  {Marchesini}, {Nelson}, {Pan}, \& {Whitaker}}]{Atek:2023}
{Atek}, H., {Chemerynska}, I., {Wang}, B., {et~al.} 2023, arXiv e-prints,
  arXiv:2305.01793, \dodoi{10.48550/arXiv.2305.01793}

\bibitem[{{Ba{\~n}ados} {et~al.}(2018){Ba{\~n}ados}, {Venemans},
  {Mazzucchelli}, {Farina}, {Walter}, {Wang}, {Decarli}, {Stern}, {Fan},
  {Davies}, {Hennawi}, {Simcoe}, {Turner}, {Rix}, {Yang}, {Kelson}, {Rudie}, \&
  {Winters}}]{Banados:2018}
{Ba{\~n}ados}, E., {Venemans}, B.~P., {Mazzucchelli}, C., {et~al.} 2018, \nat,
  553, 473, \dodoi{10.1038/nature25180}

\bibitem[{{Begelman}(2010)}]{begelman2010}
{Begelman}, M.~C. 2010, \mnras, 402, 673,
  \dodoi{10.1111/j.1365-2966.2009.15916.x}

\bibitem[{{Bezanson} {et~al.}(2022){Bezanson}, {Labbe}, {Whitaker}, {Leja},
  {Price}, {Franx}, {Brammer}, {Marchesini}, {Zitrin}, {Wang}, {Weaver},
  {Furtak}, {Atek}, {Coe}, {Cutler}, {Dayal}, {van Dokkum}, {Feldmann},
  {Forster Schreiber}, {Fujimoto}, {Geha}, {Glazebrook}, {de Graaff}, {Greene},
  {Juneau}, {Kassin}, {Kriek}, {Khullar}, {Maseda}, {Mowla}, {Muzzin},
  {Nanayakkara}, {Nelson}, {Oesch}, {Pacifici}, {Pan}, {Papovich}, {Setton},
  {Shapley}, {Smit}, {Stefanon}, {Taylor}, \& {Williams}}]{Bezanson:2022}
{Bezanson}, R., {Labbe}, I., {Whitaker}, K.~E., {et~al.} 2022, arXiv e-prints,
  arXiv:2212.04026, \dodoi{10.48550/arXiv.2212.04026}

\bibitem[{{Bogdan} {et~al.}(2023){Bogdan}, {Goulding}, {Natarajan}, {Kovacs},
  {Tremblay}, {Chadayammuri}, {Volonteri}, {Kraft}, {Forman}, {Jones},
  {Churazov}, \& {Zhuravleva}}]{Bogdan:2023}
{Bogdan}, A., {Goulding}, A., {Natarajan}, P., {et~al.} 2023, arXiv e-prints,
  arXiv:2305.15458, \dodoi{10.48550/arXiv.2305.15458}

\bibitem[{Brammer(2022)}]{Brammer:22}
Brammer, G. 2022, {msaexp: NIRSpec analyis tools}, 0.3,
  \dodoi{10.5281/zenodo.7299500}

\bibitem[{{Brammer} {et~al.}(2008){Brammer}, {van Dokkum}, \&
  {Coppi}}]{Brammer:2008}
{Brammer}, G.~B., {van Dokkum}, P.~G., \& {Coppi}, P. 2008, \apj, 686, 1503,
  \dodoi{10.1086/591786}

\bibitem[{{Bromm} \& {Loeb}(2003)}]{brommloeb2003}
{Bromm}, V., \& {Loeb}, A. 2003, \apj, 596, 34, \dodoi{10.1086/377529}

\bibitem[{{Bruzual} \& {Charlot}(2003)}]{Bruzual:2003}
{Bruzual}, G., \& {Charlot}, S. 2003, \mnras, 344, 1000,
  \dodoi{10.1046/j.1365-8711.2003.06897.x}

\bibitem[{{Buchner} {et~al.}(2014){Buchner}, {Georgakakis}, {Nandra}, {Hsu},
  {Rangel}, {Brightman}, {Merloni}, {Salvato}, {Donley}, \&
  {Kocevski}}]{Buchner:2014}
{Buchner}, J., {Georgakakis}, A., {Nandra}, K., {et~al.} 2014, \aap, 564, A125,
  \dodoi{10.1051/0004-6361/201322971}

\bibitem[{Bushouse {et~al.}(2023)Bushouse, Eisenhamer, Dencheva, Davies,
  Greenfield, Morrison, Hodge, Simon, Grumm, Droettboom, Slavich, Sosey, Pauly,
  Miller, Jedrzejewski, Hack, Davis, Crawford, Law, Gordon, Regan, Cara,
  MacDonald, Bradley, Shanahan, Jamieson, Teodoro, \&
  Williams}]{bushouse_howard_2023_7795697}
Bushouse, H., Eisenhamer, J., Dencheva, N., {et~al.} 2023, JWST Calibration
  Pipeline, 1.10.0,  Zenodo, \dodoi{10.5281/zenodo.7795697}

\bibitem[{{Carnall} {et~al.}(2019){Carnall}, {Leja}, {Johnson}, {McLure},
  {Dunlop}, \& {Conroy}}]{Carnall:2019}
{Carnall}, A.~C., {Leja}, J., {Johnson}, B.~D., {et~al.} 2019, \apj, 873, 44,
  \dodoi{10.3847/1538-4357/ab04a2}

\bibitem[{{Carnall} {et~al.}(2018){Carnall}, {McLure}, {Dunlop}, \&
  {Dav{\'e}}}]{Carnall:2018}
{Carnall}, A.~C., {McLure}, R.~J., {Dunlop}, J.~S., \& {Dav{\'e}}, R. 2018,
  \mnras, 480, 4379, \dodoi{10.1093/mnras/sty2169}

\bibitem[{{Castellano} {et~al.}(2022){Castellano}, {Fontana}, {Treu},
  {Santini}, {Merlin}, {Leethochawalit}, {Trenti}, {Vanzella}, {Mestric},
  {Bonchi}, {Belfiori}, {Nonino}, {Paris}, {Polenta}, {Roberts-Borsani},
  {Boyett}, {Brada{\v{c}}}, {Calabr{\`o}}, {Glazebrook}, {Grillo}, {Mascia},
  {Mason}, {Mercurio}, {Morishita}, {Nanayakkara}, {Pentericci}, {Rosati},
  {Vulcani}, {Wang}, \& {Yang}}]{Castellano:2022}
{Castellano}, M., {Fontana}, A., {Treu}, T., {et~al.} 2022, \apjl, 938, L15,
  \dodoi{10.3847/2041-8213/ac94d0}

\bibitem[{{Castellano} {et~al.}(2023){Castellano}, {Fontana}, {Treu}, {Merlin},
  {Santini}, {Bergamini}, {Grillo}, {Rosati}, {Acebron}, {Leethochawalit},
  {Paris}, {Bonchi}, {Belfiori}, {Calabr{\`o}}, {Correnti}, {Nonino},
  {Polenta}, {Trenti}, {Boyett}, {Brammer}, {Broadhurst}, {Caminha}, {Chen},
  {Filippenko}, {Fortuni}, {Glazebrook}, {Mascia}, {Mason}, {Menci},
  {Meneghetti}, {Mercurio}, {Metha}, {Morishita}, {Nanayakkara}, {Pentericci},
  {Roberts-Borsani}, {Roy}, {Vanzella}, {Vulcani}, {Yang}, \&
  {Wang}}]{Castellano:2023}
---. 2023, \apjl, 948, L14, \dodoi{10.3847/2041-8213/accea5}

\bibitem[{{{\c{C}}atmabacak} {et~al.}(2022){{\c{C}}atmabacak}, {Feldmann},
  {Angl{\'e}s-Alc{\'a}zar}, {Faucher-Gigu{\`e}re}, {Hopkins}, \&
  {Kere{\v{s}}}}]{Catmabacak:2022}
{{\c{C}}atmabacak}, O., {Feldmann}, R., {Angl{\'e}s-Alc{\'a}zar}, D., {et~al.}
  2022, \mnras, 511, 506, \dodoi{10.1093/mnras/stac040}

\bibitem[{{Chabrier}(2003)}]{Chabrier:2003}
{Chabrier}, G. 2003, \pasp, 115, 763, \dodoi{10.1086/376392}

\bibitem[{{Charlot} \& {Fall}(2000)}]{Charlot:2000}
{Charlot}, S., \& {Fall}, S.~M. 2000, \apj, 539, 718, \dodoi{10.1086/309250}

\bibitem[{{Curtis-Lake} {et~al.}(2023){Curtis-Lake}, {Carniani}, {Cameron},
  {Charlot}, {Jakobsen}, {Maiolino}, {Bunker}, {Witstok}, {Smit}, {Chevallard},
  {Willott}, {Ferruit}, {Arribas}, {Bonaventura}, {Curti}, {D'Eugenio},
  {Franx}, {Giardino}, {Looser}, {L{\"u}tzgendorf}, {Maseda}, {Rawle}, {Rix},
  {Rodr{\'\i}guez del Pino}, {{\"U}bler}, {Sirianni}, {Dressler}, {Egami},
  {Eisenstein}, {Endsley}, {Hainline}, {Hausen}, {Johnson}, {Rieke},
  {Robertson}, {Shivaei}, {Stark}, {Tacchella}, {Williams}, {Willmer},
  {Bhatawdekar}, {Bowler}, {Boyett}, {Chen}, {de Graaff}, {Helton}, {Hviding},
  {Jones}, {Kumari}, {Lyu}, {Nelson}, {Perna}, {Sandles}, {Saxena}, {Suess},
  {Sun}, {Topping}, {Wallace}, \& {Whitler}}]{Curtis-Lake:2023}
{Curtis-Lake}, E., {Carniani}, S., {Cameron}, A., {et~al.} 2023, Nature
  Astronomy, 7, 622, \dodoi{10.1038/s41550-023-01918-w}

\bibitem[{{Dayal} {et~al.}(2019){Dayal}, {Rossi}, {Shiralilou}, {Piana},
  {Choudhury}, \& {Volonteri}}]{Dayal:2019}
{Dayal}, P., {Rossi}, E.~M., {Shiralilou}, B., {et~al.} 2019, \mnras, 486,
  2336, \dodoi{10.1093/mnras/stz897}

\bibitem[{{Devecchi} \& {Volonteri}(2009)}]{devecchivolonteri2009}
{Devecchi}, B., \& {Volonteri}, M. 2009, \apj, 694, 302,
  \dodoi{10.1088/0004-637X/694/1/302}

\bibitem[{{Di Matteo} {et~al.}(2023){Di Matteo}, {Angles-Alcazar}, \&
  {Shankar}}]{DiMatteoetal:2023}
{Di Matteo}, T., {Angles-Alcazar}, D., \& {Shankar}, F. 2023, arXiv e-prints,
  arXiv:2304.11541, \dodoi{10.48550/arXiv.2304.11541}

\bibitem[{{Di Matteo} {et~al.}(2017){Di Matteo}, {Croft}, {Feng}, {Waters}, \&
  {Wilkins}}]{DiMatteoetal:2017}
{Di Matteo}, T., {Croft}, R. A.~C., {Feng}, Y., {Waters}, D., \& {Wilkins}, S.
  2017, \mnras, 467, 4243, \dodoi{10.1093/mnras/stx319}

\bibitem[{{Falc{\'o}n-Barroso} {et~al.}(2011){Falc{\'o}n-Barroso},
  {S{\'a}nchez-Bl{\'a}zquez}, {Vazdekis}, {Ricciardelli}, {Cardiel}, {Cenarro},
  {Gorgas}, \& {Peletier}}]{Falcon-Barroso2011}
{Falc{\'o}n-Barroso}, J., {S{\'a}nchez-Bl{\'a}zquez}, P., {Vazdekis}, A.,
  {et~al.} 2011, \aap, 532, A95, \dodoi{10.1051/0004-6361/201116842}

\bibitem[{{Fan} {et~al.}(2022){Fan}, {Banados}, \& {Simcoe}}]{Fan:2023}
{Fan}, X., {Banados}, E., \& {Simcoe}, R.~A. 2022, arXiv e-prints,
  arXiv:2212.06907, \dodoi{10.48550/arXiv.2212.06907}

\bibitem[{{Fan} {et~al.}(2019){Fan}, {Wang}, {Yang}, {Keeton}, {Yue},
  {Zabludoff}, {Bian}, {Bonaglia}, {Georgiev}, {Hennawi}, {Li}, {McGreer},
  {Naidu}, {Pacucci}, {Rabien}, {Thompson}, {Venemans}, {Walter}, {Wang}, \&
  {Wu}}]{Fan:2019}
{Fan}, X., {Wang}, F., {Yang}, J., {et~al.} 2019, \apjl, 870, L11,
  \dodoi{10.3847/2041-8213/aaeffe}

\bibitem[{{Ferland} {et~al.}(2017){Ferland}, {Chatzikos}, {Guzm{\'a}n},
  {Lykins}, {van Hoof}, {Williams}, {Abel}, {Badnell}, {Keenan}, {Porter}, \&
  {Stancil}}]{Ferland:2017}
{Ferland}, G.~J., {Chatzikos}, M., {Guzm{\'a}n}, F., {et~al.} 2017, \rmxaa, 53,
  385, \dodoi{10.48550/arXiv.1705.10877}

\bibitem[{{Feroz} {et~al.}(2019){Feroz}, {Hobson}, {Cameron}, \&
  {Pettitt}}]{Feroz:2019}
{Feroz}, F., {Hobson}, M.~P., {Cameron}, E., \& {Pettitt}, A.~N. 2019, The Open
  Journal of Astrophysics, 2, 10, \dodoi{10.21105/astro.1306.2144}

\bibitem[{{Freitag} {et~al.}(2006){Freitag}, {G{\"u}rkan}, \&
  {Rasio}}]{freitagetal2006}
{Freitag}, M., {G{\"u}rkan}, M.~A., \& {Rasio}, F.~A. 2006, \mnras, 368, 141,
  \dodoi{10.1111/j.1365-2966.2006.10096.x}

\bibitem[{{Furtak} {et~al.}(2023{\natexlab{a}}){Furtak}, {Zitrin}, {Plat},
  {Fujimoto}, {Wang}, {Nelson}, {Labb{\'e}}, {Bezanson}, {Brammer}, {van
  Dokkum}, {Endsley}, {Glazebrook}, {Greene}, {Leja}, {Price}, {Smit}, {Stark},
  {Weaver}, {Whitaker}, {Atek}, {Chevallard}, {Curtis-Lake}, {Dayal}, {Feltre},
  {Franx}, {Fudamoto}, {Marchesini}, {Mowla}, {Pan}, {Suess},
  {Vidal-Garc{\'\i}a}, \& {Williams}}]{Furtak:2022}
{Furtak}, L.~J., {Zitrin}, A., {Plat}, A., {et~al.} 2023{\natexlab{a}}, \apj,
  952, 142, \dodoi{10.3847/1538-4357/acdc9d}

\bibitem[{{Furtak} {et~al.}(2023{\natexlab{b}}){Furtak}, {Zitrin}, {Weaver},
  {Atek}, {Bezanson}, {Labb{\'e}}, {Whitaker}, {Leja}, {Price}, {Brammer},
  {Wang}, {Marchesini}, {Pan}, {Dayal}, {van Dokkum}, {Feldmann}, {Fujimoto},
  {Franx}, {Khullar}, {Nelson}, \& {Mowla}}]{Furtak:2023b}
{Furtak}, L.~J., {Zitrin}, A., {Weaver}, J.~R., {et~al.} 2023{\natexlab{b}},
  \mnras, 523, 4568, \dodoi{10.1093/mnras/stad1627}

\bibitem[{{Furtak} {et~al.}(2023{\natexlab{c}}){Furtak}, {Labb{\'e}}, {Zitrin},
  {Greene}, {Dayal}, {Chemerynska}, {Kokorev}, {Miller}, {Goulding},
  {Bezanson}, {Brammer}, {Cutler}, {Leja}, {Pan}, {Price}, {Wang}, {Weaver},
  {Whitaker}, {Atek}, {Bogd{\'a}n}, {Charlot}, {Curtis-Lake}, {van Dokkum},
  {Endsley}, {Fudamoto}, {Fujimoto}, {de Graaff}, {Glazebrook}, {Juneau},
  {Marchesini}, {Maseda}, {Nelson}, {Oesch}, {Plat}, {Setton}, {Stark}, \&
  {Williams}}]{Furtak_nature:2023}
{Furtak}, L.~J., {Labb{\'e}}, I., {Zitrin}, A., {et~al.} 2023{\natexlab{c}},
  arXiv e-prints, arXiv:2308.05735, \dodoi{10.48550/arXiv.2308.05735}

\bibitem[{{Goulding} \& {Alexander}(2009)}]{Goulding:2009}
{Goulding}, A.~D., \& {Alexander}, D.~M. 2009, \mnras, 398, 1165,
  \dodoi{10.1111/j.1365-2966.2009.15194.x}

\bibitem[{{Greene} {et~al.}(2020){Greene}, {Strader}, \&
  {Ho}}]{Greeneetal:2020}
{Greene}, J.~E., {Strader}, J., \& {Ho}, L.~C. 2020, \araa, 58, 257,
  \dodoi{10.1146/annurev-astro-032620-021835}

\bibitem[{{Habouzit} {et~al.}(2022){Habouzit}, {Onoue}, {Ba{\~n}ados},
  {Neeleman}, {Angl{\'e}s-Alc{\'a}zar}, {Walter}, {Pillepich}, {Dav{\'e}},
  {Jahnke}, \& {Dubois}}]{Habouzit:2022}
{Habouzit}, M., {Onoue}, M., {Ba{\~n}ados}, E., {et~al.} 2022, \mnras, 511,
  3751, \dodoi{10.1093/mnras/stac225}

\bibitem[{{Haiman} \& {Loeb}(2001)}]{Haiman:2001}
{Haiman}, Z., \& {Loeb}, A. 2001, \apj, 552, 459, \dodoi{10.1086/320586}

\bibitem[{{Harikane} {et~al.}(2022){Harikane}, {Ono}, {Ouchi}, {Liu},
  {Sawicki}, {Shibuya}, {Behroozi}, {He}, {Shimasaku}, {Arnouts}, {Coupon},
  {Fujimoto}, {Gwyn}, {Huang}, {Inoue}, {Kashikawa}, {Komiyama}, {Matsuoka}, \&
  {Willott}}]{Harikane:2022}
{Harikane}, Y., {Ono}, Y., {Ouchi}, M., {et~al.} 2022, \apjs, 259, 20,
  \dodoi{10.3847/1538-4365/ac3dfc}

\bibitem[{{Harikane} {et~al.}(2023){Harikane}, {Zhang}, {Nakajima}, {Ouchi},
  {Isobe}, {Ono}, {Hatano}, {Xu}, \& {Umeda}}]{Harikane:2023}
{Harikane}, Y., {Zhang}, Y., {Nakajima}, K., {et~al.} 2023, arXiv e-prints,
  arXiv:2303.11946, \dodoi{10.48550/arXiv.2303.11946}

\bibitem[{Hoffman {et~al.}(2014)Hoffman, Gelman, {et~al.}}]{Hoffman2014}
Hoffman, M.~D., Gelman, A., {et~al.} 2014, J. Mach. Learn. Res., 15, 1593

\bibitem[{{Horne}(1986)}]{Horne:86}
{Horne}, K. 1986, \pasp, 98, 609, \dodoi{10.1086/131801}

\bibitem[{{Inayoshi} {et~al.}(2022){Inayoshi}, {Onoue}, {Sugahara}, {Inoue}, \&
  {Ho}}]{Inayoshi:2022}
{Inayoshi}, K., {Onoue}, M., {Sugahara}, Y., {Inoue}, A.~K., \& {Ho}, L.~C.
  2022, \apjl, 931, L25, \dodoi{10.3847/2041-8213/ac6f01}

\bibitem[{{Izumi} {et~al.}(2019){Izumi}, {Onoue}, {Matsuoka}, {Nagao},
  {Strauss}, {Imanishi}, {Kashikawa}, {Fujimoto}, {Kohno}, {Toba}, {Umehata},
  {Goto}, {Ueda}, {Shirakata}, {Silverman}, {Greene}, {Harikane}, {Hashimoto},
  {Ikarashi}, {Iono}, {Iwasawa}, {Lee}, {Minezaki}, {Nakanishi}, {Tamura},
  {Tang}, \& {Taniguchi}}]{Izumi:2019}
{Izumi}, T., {Onoue}, M., {Matsuoka}, Y., {et~al.} 2019, \pasj, 71, 111,
  \dodoi{10.1093/pasj/psz096}

\bibitem[{{Johnson} {et~al.}(2021){Johnson}, {Leja}, {Conroy}, \&
  {Speagle}}]{Johnson:2021}
{Johnson}, B.~D., {Leja}, J., {Conroy}, C., \& {Speagle}, J.~S. 2021, \apjs,
  254, 22, \dodoi{10.3847/1538-4365/abef67}

\bibitem[{{Kocevski} {et~al.}(2023){Kocevski}, {Onoue}, {Inayoshi}, {Trump},
  {Arrabal Haro}, {Grazian}, {Dickinson}, {Finkelstein}, {Kartaltepe},
  {Hirschmann}, {Fujimoto}, {Juneau}, {Amorin}, {Bagley}, {Barro}, {Bell},
  {Bisigello}, {Calabro}, {Cleri}, {Cooper}, {Ding}, {Grogin}, {Ho}, {Inoue},
  {Jiang}, {Jones}, {Koekemoer}, {Li}, {Li}, {McGrath}, {Molina}, {Papovich},
  {Perez-Gonzalez}, {Pirzkal}, {Wilkins}, {Yang}, \& {Yung}}]{Kocevski:2023}
{Kocevski}, D.~D., {Onoue}, M., {Inayoshi}, K., {et~al.} 2023, arXiv e-prints,
  arXiv:2302.00012, \dodoi{10.48550/arXiv.2302.00012}

\bibitem[{{Kormendy} \& {Ho}(2013)}]{kormendyho2013}
{Kormendy}, J., \& {Ho}, L.~C. 2013, \araa, 51, 511,
  \dodoi{10.1146/annurev-astro-082708-101811}

\bibitem[{{Koushiappas} {et~al.}(2004){Koushiappas}, {Bullock}, \&
  {Dekel}}]{koushiappasetal2004}
{Koushiappas}, S.~M., {Bullock}, J.~S., \& {Dekel}, A. 2004, \mnras, 354, 292,
  \dodoi{10.1111/j.1365-2966.2004.08190.x}

\bibitem[{{Labbe} {et~al.}(2023){Labbe}, {Greene}, {Bezanson}, {Fujimoto},
  {Furtak}, {Goulding}, {Matthee}, {Naidu}, {Oesch}, {Atek}, {Brammer},
  {Chemerynska}, {Coe}, {Cutler}, {Dayal}, {Feldmann}, {Franx}, {Glazebrook},
  {Leja}, {Marchesini}, {Maseda}, {Nanayakkara}, {Nelson}, {Pan}, {Papovich},
  {Price}, {Suess}, {Wang}, {Whitaker}, {Williams}, \&
  {Zitrin}}]{LabbeGreene:2023}
{Labbe}, I., {Greene}, J.~E., {Bezanson}, R., {et~al.} 2023, arXiv e-prints,
  arXiv:2306.07320, \dodoi{10.48550/arXiv.2306.07320}

\bibitem[{{Larson} {et~al.}(2023){Larson}, {Finkelstein}, {Kocevski},
  {Hutchison}, {Trump}, {Arrabal Haro}, {Bromm}, {Cleri}, {Dickinson},
  {Fujimoto}, {Kartaltepe}, {Koekemoer}, {Papovich}, {Pirzkal}, {Tacchella},
  {Zavala}, {Bagley}, {Behroozi}, {Champagne}, {Cole}, {Jung}, {Morales},
  {Yang}, {Zhang}, {Zitrin}, {Amor{\'\i}n}, {Burgarella}, {Casey}, {Ch{\'a}vez
  Ortiz}, {Cox}, {Chworowsky}, {Fontana}, {Gawiser}, {Grazian}, {Grogin},
  {Harish}, {Hathi}, {Hirschmann}, {Holwerda}, {Juneau}, {Leung}, {Lucas},
  {McGrath}, {P{\'e}rez-Gonz{\'a}lez}, {Rigby}, {Seill{\'e}}, {Simons},
  {Weiner}, {Wilkins}, {Yung}, \& {The CEERS Team}}]{Larson:2023}
{Larson}, R.~L., {Finkelstein}, S.~L., {Kocevski}, D.~D., {et~al.} 2023, arXiv
  e-prints, arXiv:2303.08918, \dodoi{10.48550/arXiv.2303.08918}

\bibitem[{{Lodato} \& {Natarajan}(2006)}]{lodatonatarajan2006}
{Lodato}, G., \& {Natarajan}, P. 2006, \mnras, 371, 1813,
  \dodoi{10.1111/j.1365-2966.2006.10801.x}

\bibitem[{{Lodato} \& {Natarajan}(2007)}]{LodatoPN:2007}
---. 2007, \mnras, 377, L64, \dodoi{10.1111/j.1745-3933.2007.00304.x}

\bibitem[{{Loeb} \& {Rasio}(1994)}]{loebrasio1994}
{Loeb}, A., \& {Rasio}, F.~A. 1994, \apj, 432, 52, \dodoi{10.1086/174548}

\bibitem[{{Maiolino} {et~al.}(2023{\natexlab{a}}){Maiolino}, {Scholtz},
  {Witstok}, {Carniani}, {D'Eugenio}, {de Graaff}, {Uebler}, {Tacchella},
  {Curtis-Lake}, {Arribas}, {Bunker}, {Charlot}, {Chevallard}, {Curti},
  {Looser}, {Maseda}, {Rawle}, {Rodriguez Del Pino}, {Willott}, {Egami},
  {Eisenstein}, {Hainline}, {Robertson}, {Williams}, {Willmer}, {Baker},
  {Boyett}, {DeCoursey}, {Fabian}, {Helton}, {Ji}, {Jones}, {Kumari},
  {Laporte}, {Nelson}, {Perna}, {Sandles}, {Shivaei}, \&
  {Sun}}]{Maiolino:2023a}
{Maiolino}, R., {Scholtz}, J., {Witstok}, J., {et~al.} 2023{\natexlab{a}},
  arXiv e-prints, arXiv:2305.12492, \dodoi{10.48550/arXiv.2305.12492}

\bibitem[{{Maiolino} {et~al.}(2023{\natexlab{b}}){Maiolino}, {Scholtz},
  {Curtis-Lake}, {Carniani}, {Baker}, {de Graaff}, {Tacchella}, {{\"U}bler},
  {D'Eugenio}, {Witstok}, {Curti}, {Arribas}, {Bunker}, {Charlot},
  {Chevallard}, {Eisenstein}, {Egami}, {Ji}, {Jones}, {Lyu}, {Rawle},
  {Robertson}, {Rujopakarn}, {Perna}, {Sun}, {Venturi}, {Williams}, \&
  {Willott}}]{Maiolino:2023b}
{Maiolino}, R., {Scholtz}, J., {Curtis-Lake}, E., {et~al.} 2023{\natexlab{b}},
  arXiv e-prints, arXiv:2308.01230, \dodoi{10.48550/arXiv.2308.01230}

\bibitem[{{Matsuoka} {et~al.}(2018){Matsuoka}, {Strauss}, {Kashikawa}, {Onoue},
  {Iwasawa}, {Tang}, {Lee}, {Imanishi}, {Nagao}, {Akiyama}, {Asami}, {Bosch},
  {Furusawa}, {Goto}, {Gunn}, {Harikane}, {Ikeda}, {Izumi}, {Kawaguchi},
  {Kato}, {Kikuta}, {Kohno}, {Komiyama}, {Lupton}, {Minezaki}, {Miyazaki},
  {Murayama}, {Niida}, {Nishizawa}, {Noboriguchi}, {Oguri}, {Ono}, {Ouchi},
  {Price}, {Sameshima}, {Schulze}, {Shirakata}, {Silverman}, {Sugiyama},
  {Tait}, {Takada}, {Takata}, {Tanaka}, {Toba}, {Utsumi}, {Wang}, \&
  {Yamashita}}]{Matsuoka:2018}
{Matsuoka}, Y., {Strauss}, M.~A., {Kashikawa}, N., {et~al.} 2018, \apj, 869,
  150, \dodoi{10.3847/1538-4357/aaee7a}

\bibitem[{{Matsuoka} {et~al.}(2023){Matsuoka}, {Onoue}, {Iwasawa}, {Strauss},
  {Kashikawa}, {Izumi}, {Nagao}, {Imanishi}, {Akiyama}, {Silverman}, {Asami},
  {Bosch}, {Furusawa}, {Goto}, {Gunn}, {Harikane}, {Ikeda}, {Inayoshi},
  {Ishimoto}, {Kawaguchi}, {Kikuta}, {Kohno}, {Komiyama}, {Lee}, {Lupton},
  {Minezaki}, {Miyazaki}, {Murayama}, {Nishizawa}, {Oguri}, {Ono}, {Oogi},
  {Ouchi}, {Price}, {Sameshima}, {Sugiyama}, {Tait}, {Takada}, {Takahashi},
  {Takata}, {Tanaka}, {Toba}, {Wang}, \& {Yamashita}}]{Matsuoka:2023}
{Matsuoka}, Y., {Onoue}, M., {Iwasawa}, K., {et~al.} 2023, arXiv e-prints,
  arXiv:2305.11225, \dodoi{10.48550/arXiv.2305.11225}

\bibitem[{{Matt} {et~al.}(2000){Matt}, {Fabian}, {Guainazzi}, {Iwasawa},
  {Bassani}, \& {Malaguti}}]{Matt:2000}
{Matt}, G., {Fabian}, A.~C., {Guainazzi}, M., {et~al.} 2000, \mnras, 318, 173,
  \dodoi{10.1046/j.1365-8711.2000.03721.x}

\bibitem[{{Matthee} {et~al.}(2023){Matthee}, {Naidu}, {Brammer}, {Chisholm},
  {Eilers}, {Goulding}, {Greene}, {Kashino}, {Labbe}, {Lilly}, {Mackenzie},
  {Oesch}, {Weibel}, {Wuyts}, {Xiao}, {Bordoloi}, {Bouwens}, {van Dokkum},
  {Illingworth}, {Kramarenko}, {Maseda}, {Mason}, {Meyer}, {Nelson}, {Reddy},
  {Shivaei}, {Simcoe}, \& {Yue}}]{Matthee:2023}
{Matthee}, J., {Naidu}, R.~P., {Brammer}, G., {et~al.} 2023, arXiv e-prints,
  arXiv:2306.05448, \dodoi{10.48550/arXiv.2306.05448}

\bibitem[{{Miller} \& {Hamilton}(2002)}]{millerhamilton2002}
{Miller}, M.~C., \& {Hamilton}, D.~P. 2002, \mnras, 330, 232,
  \dodoi{10.1046/j.1365-8711.2002.05112.x}

\bibitem[{{Mortlock} {et~al.}(2011){Mortlock}, {Warren}, {Venemans}, {Patel},
  {Hewett}, {McMahon}, {Simpson}, {Theuns}, {Gonz{\'a}les-Solares}, {Adamson},
  {Dye}, {Hambly}, {Hirst}, {Irwin}, {Kuiper}, {Lawrence}, \&
  {R{\"o}ttgering}}]{Mortlock:2011}
{Mortlock}, D.~J., {Warren}, S.~J., {Venemans}, B.~P., {et~al.} 2011, \nat,
  474, 616, \dodoi{10.1038/nature10159}

\bibitem[{{Natarajan}(2011)}]{Natarajan:2011}
{Natarajan}, P. 2011, Bulletin of the Astronomical Society of India, 39, 145,
  \dodoi{10.48550/arXiv.1104.4797}

\bibitem[{{Natarajan}(2021)}]{Natarajan:2021}
---. 2021, \mnras, 501, 1413, \dodoi{10.1093/mnras/staa3724}

\bibitem[{{Natarajan} {et~al.}(2017){Natarajan}, {Pacucci}, {Ferrara},
  {Agarwal}, {Ricarte}, {Zackrisson}, \& {Cappelluti}}]{natarajanetal2017}
{Natarajan}, P., {Pacucci}, F., {Ferrara}, A., {et~al.} 2017, The Astrophysical
  Journal, 838, 117, \dodoi{10.3847/1538-4357/aa6330}

\bibitem[{{Natarajan} {et~al.}(2023){Natarajan}, {Pacucci}, {Ricarte},
  {Bogdan}, {Goulding}, \& {Cappelluti}}]{Natarajan:2023}
{Natarajan}, P., {Pacucci}, F., {Ricarte}, A., {et~al.} 2023, arXiv e-prints,
  arXiv:2308.02654.
\newblock \doarXiv{2308.02654}

\bibitem[{{Neeleman} {et~al.}(2021){Neeleman}, {Novak}, {Venemans}, {Walter},
  {Decarli}, {Kaasinen}, {Schindler}, {Ba{\~n}ados}, {Carilli}, {Drake}, {Fan},
  \& {Rix}}]{Neeleman:2021}
{Neeleman}, M., {Novak}, M., {Venemans}, B.~P., {et~al.} 2021, \apj, 911, 141,
  \dodoi{10.3847/1538-4357/abe70f}

\bibitem[{{Pasha} \& {Miller}(2023)}]{Pasha:2023}
{Pasha}, I., \& {Miller}, T.~B. 2023, arXiv e-prints, arXiv:2306.05454,
  \dodoi{10.48550/arXiv.2306.05454}

\bibitem[{Phan {et~al.}(2019)Phan, Pradhan, \& Jankowiak}]{Phan2019}
Phan, D., Pradhan, N., \& Jankowiak, M. 2019, arXiv preprint arXiv:1912.11554

\bibitem[{{Portegies Zwart} \& {McMillan}(2002)}]{portegieszwartetal2002}
{Portegies Zwart}, S.~F., \& {McMillan}, S.~L.~W. 2002, \apj, 576, 899,
  \dodoi{10.1086/341798}

\bibitem[{{Reines} \& {Volonteri}(2015)}]{reinesvolonteri2015}
{Reines}, A.~E., \& {Volonteri}, M. 2015, \apj, 813, 82,
  \dodoi{10.1088/0004-637X/813/2/82}

\bibitem[{{S{\'a}nchez-Bl{\'a}zquez} {et~al.}(2006){S{\'a}nchez-Bl{\'a}zquez},
  {Peletier}, {Jim{\'e}nez-Vicente}, {Cardiel}, {Cenarro},
  {Falc{\'o}n-Barroso}, {Gorgas}, {Selam}, \&
  {Vazdekis}}]{Sanchez-Bazquez:2006}
{S{\'a}nchez-Bl{\'a}zquez}, P., {Peletier}, R.~F., {Jim{\'e}nez-Vicente}, J.,
  {et~al.} 2006, \mnras, 371, 703, \dodoi{10.1111/j.1365-2966.2006.10699.x}

\bibitem[{{Volonteri} {et~al.}(2008){Volonteri}, {Lodato}, \&
  {Natarajan}}]{VolonteriGLPN:2008}
{Volonteri}, M., {Lodato}, G., \& {Natarajan}, P. 2008, \mnras, 383, 1079,
  \dodoi{10.1111/j.1365-2966.2007.12589.x}

\bibitem[{{Wang} {et~al.}(2023){Wang}, {Leja}, {Bezanson}, {Johnson},
  {Khullar}, {Labb{\'e}}, {Price}, {Weaver}, \& {Whitaker}}]{Wang:2023}
{Wang}, B., {Leja}, J., {Bezanson}, R., {et~al.} 2023, \apjl, 944, L58,
  \dodoi{10.3847/2041-8213/acba99}

\bibitem[{{Weaver} {et~al.}(2023){Weaver}, {Cutler}, {Pan}, {Whitaker},
  {Labbe}, {Price}, {Bezanson}, {Brammer}, {Marchesini}, {Leja}, {Wang},
  {Furtak}, {Zitrin}, {Atek}, {Coe}, {Dayal}, {van Dokkum}, {Feldmann},
  {Forster Schreiber}, {Franx}, {Fujimoto}, {Fudamoto}, {Glazebrook}, {de
  Graaff}, {Greene}, {Juneau}, {Kassin}, {Kriek}, {Khullar}, {Maseda}, {Mowla},
  {Muzzin}, {Nanayakkara}, {Nelson}, {Oesch}, {Pacifici}, {Papovich}, {Setton},
  {Shapley}, {Smit}, {Stefanon}, {Taylor}, {Weibel}, \&
  {Williams}}]{Weaver:2023}
{Weaver}, J.~R., {Cutler}, S.~E., {Pan}, R., {et~al.} 2023, arXiv e-prints,
  arXiv:2301.02671, \dodoi{10.48550/arXiv.2301.02671}

\bibitem[{{Yaqoob}(2012)}]{Yaqoob:2012}
{Yaqoob}, T. 2012, \mnras, 423, 3360, \dodoi{10.1111/j.1365-2966.2012.21129.x}

\end{thebibliography}

\end{document}